\documentstyle[12pt]{article}

\begin{document}
 \thispagestyle{empty}

 \def\ve#1{\vert #1\rangle}
 \def\vc#1{\langle #1\vert}

 \newcommand{\p}[1]{(\ref{#1})}
 \newcommand{\be}{\begin{equation}}
 \newcommand{\ee}{\end{equation}}
 \newcommand{\sect}[1]{\setcounter{equation}{0}\section{#1}}

 \renewcommand{\theequation}{\thesection.\arabic{equation}}

 \newcommand{\vs}[1]{\rule[- #1 mm]{0mm}{#1 mm}}
 \newcommand{\hs}[1]{\hspace{#1mm}}
 \newcommand{\mb}[1]{\hs{5}\mbox{#1}\hs{5}}
 \newcommand{\Db}{{\overline D}}
 \newcommand{\bea}{\begin{eqnarray}}
 \newcommand{\eea}{\end{eqnarray}}
 \newcommand{\wt}[1]{\widetilde{#1}}
 \newcommand{\und}[1]{\underline{#1}}
 \newcommand{\ov}[1]{\overline{#1}}
 \newcommand{\sm}[2]{\frac{\mbox{\footnotesize #1}\vs{-2}}
            {\vs{-2}\mbox{\footnotesize #2}}}
 \newcommand{\prt}{\partial}
 \newcommand{\eps}{\epsilon}

 \newcommand{\R}{\mbox{\rule{0.2mm}{2.8mm}\hspace{-1.5mm} R}}
 \newcommand{\Z}{Z\hspace{-2mm}Z}

 \newcommand{\cd}{{\cal D}}
 \newcommand{\cg}{{\cal G}}
 \newcommand{\ck}{{\cal K}}
 \newcommand{\cw}{{\cal W}}

 \newcommand{\vj}{\vec{J}}
 \newcommand{\vl}{\vec{\lambda}}
 \newcommand{\vz}{\vec{\sigma}}
 \newcommand{\vt}{\vec{\tau}}
 \newcommand{\vw}{\vec{W}}
 \newcommand{\poiss}{\stackrel{\otimes}{,}}

 \def\l#1#2{\raisebox{.2ex}{$\displaystyle
   \mathop{#1}^{{\scriptstyle #2}\rightarrow}$}}
 \def\r#1#2{\raisebox{.2ex}{$\displaystyle
  \mathop{#1}^{\leftarrow {\scriptstyle #2}}$}}

 \renewcommand{\thefootnote}{\fnsymbol{footnote}}

 \setcounter{page}{1}

 \pagestyle{empty}
 \newpage


 \vs{8}

 \begin{center}
 {\LARGE {\bf Representation of the quantum algebra
 $SU_q(2)$}}\\[0.6cm]
 {\LARGE {\bf in the basis with diagonal $"J_x"$ generator}}\\[1cm]

 \vs{8}

 {\large A.N. Leznov$^{b,c}$}
 {}~\\
 \quad \\

 {\it {$^{(b)}$ Institute for High Energy Physics,}}\\
   {\em 142284 Protvino, Moscow Region, Russia}\\
 {\em {~$~^{(c)}$ Bogoliubov Laboratory of Theoretical Physics,
 JINR,}}\\
 {\em 141980 Dubna, Moscow Region, Russia}

 \end{center}

 \bigskip
 \bigskip
 \bigskip


 \begin{abstract}

 Generators of the quantum $SU_q(2)$ algebra are obtained in the
 explicit form in the basis where the operator
 $\exp {J_z\over 2} J_x \exp {J_z\over 2}$ is diagonal.
 It is shown that the solution of this problem is related to the
 representation theory of the two-dimensional algebra
 $[s,r]=\tanh t\, (s^2-r^2+1)$.
 The relevance of such basis to some problems of quantum optics is
 discussed.

 \end{abstract}

 \vfill

{\em E-Mail:\
leznov@ce.ifisicam.unam.mx}

\pagestyle{plain}

\renewcommand{\thefootnote}{\arabic{footnote}}

\setcounter{footnote}{0}

\section{Introduction}

The common realizations of the quantum algebra SU${}_q$(2) are
connected
with the basis where the generator $J_z$ is diagonal. This tradition
can be
explained by the fact that this algebra is invariant under the
U(1)-rotations in $x,y$ plane. However, in the recent paper \cite{1}
it was shown that in each irreducible representation of SU${}_q$(2)
algebra
the operator
$\exp {J_z t\over 2} J_x \exp {J_z t\over 2}$ can be diagonalized, and
its
spectrum does not depend on the choice of the representation.

The aim of the present paper is to show that this result is not
occasional
but has deep foundations. We will be able to construct in the explicit
form
generators of irreducible representations of $SU_q(2)$ algebra in the
above
mentioned basis. The strategy of our calculations is based upon the
quasi-classical nature of representation of the quantum algebras, as
it was
observed in \cite{ZET}. Firstly, we will find the representation of
the corresponding functional quantum group and then
generalize these results properly to the quantum algebra case.

The paper is organized in the following way. In Section 2 we consider
the 
basis in quantum algebra $SU_q(2)$, where operator $"J_x"$ is
diagonal. We
show that the $SU_q(2)$ generators can be written in terms of only two
operators
from the enveloping algebra (see $S$ and $r$ below, $s$ is
proportional to
$"J_x"$ generator), which obey a single commutation relations. In
Sections 3 
and 4 we find realization of this new $\{s,r\}$ algebrain terms of
generators of 
Heisenberg-Weyl algebra. We start with the consideration at the
classical
level, when all commutators are changed for the Poisson brackets ( a
functional 
group case) in Section 3. In Section 4 we give solution for the
"quantum"
case. The limit, when deformation parameter $t\to 0,(q\to 1)$ is
considered in 
Section 5. The raising and lowering operators in the new basis are
found
in Sectoion 6. Nontrivial role of the Casimir operator is discussed in
Section
7, where finite-dimensional representations of $\{s,r\}$ algebra are 
represented in explicit form. Conclusions are given in Section 8.

\section{New form of commutation relations of the SU${}_q$(2) algebra}

The common form of the commutation relations for the generators
$ J^{\pm},\, H$ of the SU${}_q$(2) algebra is :
\begin{equation}
\label{1}
  [H,J^{\pm}]=\pm 2 J^{\pm},\quad [J^+,J^-]={\sinh (tH)\over \sinh t},
  \quad
    J_x=J^++J^-,\quad H=2J_z,
\end{equation}
where $t=\log q$ is the deformation parameter. The Casimir operator
that
commutes with all generators of the algebra has the form:
$$
 {\cal C}=J^+J^-+J^-J^+ +{\cosh t\over \sinh^2 t}\,\left(\cosh
 (tH)-1\right)
$$
Let us introduce the generators
$$
T^{\pm}=\exp {Ht\over 4}\, J^{\pm} \exp {Ht\over 4},\qquad
R=\exp(tH)\,.
$$
The commutation relations determining the SU${}_q$(2)
algebra (\ref{1}) may be rewritten as follows:
\begin{equation}
  RT^{\pm}= e^{\pm 2t}\, T^{\pm}R,\qquad
  e^t\, T^+T^--e^{-t}\, T^-T^+ ={R^2-1\over  2\sinh t}\,\label{2}.
\end{equation}
The Casimir operator takes the form:
$$
  {\cal C}=R^{-1}\left(
    e^t\, T^+T^-+e^{-t}\, T^-T^+ + {(R-1)^2 \cosh t\over 2\sinh^2 t}
   \right)
$$
The remarkable property of the commutation relations (\ref{2})
is that the last equation may be rewritten in terms of only two
(not three!) generators of the initial algebra.
Introducing the pair of operators
$$
  Q^{\pm}=T^{\pm}\pm {R\over 2\sinh t},
$$
we rewrite the last equation of (\ref{2}) in the form:
\begin{equation}
  e^t\, Q^+Q^--e^{-t}\, Q^-Q^+=-{1\over 2\sinh t}\,\label{3}.
\end{equation}
Note that
$$
Q^++Q^-=T^++T^-=\exp {J_z t\over 2} J_x \exp {J_z t\over 2}, \quad
H=2J_z
$$
is exactly the operator which was diagonalized in \cite{1}.

Now we want to consider the representations of SU${}_q$(2) algebra
in the basis where this operator is diagonal. Our strategy
is as follows: We will try to find solutions of the single equation
(\ref{3}), and express $R$ in terms
of $ Q^+,Q^-$ and the Casimir operator.
We start with the less difficult second part of the problem.

Let us introduce the generators
$$
  s=(Q^++Q^-)\sinh t,\qquad  r=(Q^+-Q^-)\sinh t\,.
$$
We now can rewrite (\ref{3}) in the more attractive form:
\begin{equation}
  [s,r]=(s^2-r^2+1)\tanh t \,.  \label{4}
\end{equation}

It is clear that the two-dimensional quantum algebra (\ref{3}) is
invariant
with respect to $U(1)$ transformations $Q^{\pm}\to \exp \pm \alpha
Q^{\pm}$. As a consequence, the algebra (\ref{4}) is invariant with
respect
to the Lorenz transformations in the two-dimensional $r,s$ plane.

By means of purely algebraic manipulations using the above
formulae, it is not difficult to rewrite the Casimir operator
in terms of $s,r,R$ operators:
$$
  {\cal C}=\sinh^{-2} t
   R^{-1}\, \left[ \frac{s^2-r^2+1}{2\cosh t} + R \left((r-1)\,\cosh t
     - s \,\sinh t\right) \right]
$$
The remarkable fact is that the quadratic terms $(R^2)$ in the square
brackets in the right-hand side of the last expression are
canceled and this give the possibility to express $R$ in terms of
$s,r$
operators under the fixed numerical value of the Casimir operator,
\begin{equation}
  R= K^{-1}\, (s^2-r^2+1)/2\,, \quad
  K = \cosh t\,\left({\cal C} \sinh^2 t
      + (1-r)\,\cosh t - s \,\sinh t\right)\,\label{5},
\end{equation}
provided that the inverse of the operator $K$ exists.
We recall that the operators $s,r$ do not commute and thus the
order of the factors in the above formulae is important.

In conclusion of this section we describe the way of proving that
the operator $R$ as defined by (\ref{5}), indeed satisfies
the first pair of equations of initial quantum algebra (\ref{2}).
We restrict ourselves by $+$ sign in Eq.\ (\ref{2}).

In accordance with the above definitions,
$$
     T^{\pm} = \frac{s\pm r\mp R}{2\sinh t}\,.
$$
Thus, for the checking of (\ref{2}) it is necessary
to move $s+r$ throughout $R$. The following formulae explain this
procedure:
$$
 (s-r)(s+r)+1 =s^2-r^2+[s,r]+1={\exp t\over \cosh t}\,(s^2-r^2+1)
$$
$$
 (s+r)(s-r)+1 =s^2-r^2-[s,r]+1={\exp(-t) \over \cosh t}\,(s^2-r^2+1)
$$
$$
  (s^2-r^2+1)\,(s+r)= e^t \cosh t\, ((s+r)(s-r)+1)(s+r)=
$$
$$
  e^t \cosh t\, (s+r)((s-r)(s+r)+1)= e^{2t}(s+r)(s^2-r^2+1)\,.
$$
These formulae explain the action connected with the second factor
of $R=K^{-1} (s^2-r^2+1)/2$. The calculations connected with the first
factor
are based upon the commutator:
$$
[K,\,s+r]=-4 e^{-t} \cosh t\,[s,r] = -2 (1- e^{-2t})(s^2-r^2+1)\,.
$$
Multiplying the last equations from the left and from the right
by $K^{-1}$ we come to the formula
$$
  [s+r,K^{-1}]=-2(1-e^{-2t})\,K^{-1}(s^2-r^2+1)K^{-1}
$$
which completes checking of Eq.\ (\ref{2}).

\section{Resolution of the two-dimensional algebra on the
level of the functional group}

On the functional group level  \cite{Eisenhart}
commutator in (\ref{4})
is replaced by Poisson brackets. We thus have,
\begin{equation}
\label{4F}
  \{s,r\}=\tanh t\,(s^2-r^2+1)\,.
\end{equation}
In accordance with the Darboux's theorem \cite{Eisenhart}, $\,s,r$
from
(\ref{4F})  may  be  expressed  as   functions  of  the  pair   of
canonically conjugated variables $p,x,\, \{p,x\}=1$.

To keep the correspondence with the results \cite{1} let us choose
$s=\sinh tp$ and $r=r(x,p)$.  Eq.\ (\ref{4F}) takes on the form
$$
  t\cosh(tp) \,\frac{dr}{dx}=\tanh t\,(\cosh^2 tp - r^2)
$$
with the obvious general solution:
$$
  r=\cosh(tp)\, \coth \left( \nu x + \phi(p)\right),
  \qquad \nu\equiv {\tanh t\over t}\,.
$$

\section{The case of the quantum algebra}

Reminding about the quasi-classical nature of the problem of
construction
of irreducible representation of the quantum algebras \cite{ZET},
and keeping in mind the results of the consideration on the functional
group
level (the previous section), let us try to resolve the quantum
algebra
(\ref{4}) with the help of the substitution:
\begin{equation}
  s=\sinh (tp)\,,\qquad
  r=\frac{A(x)(x) \,e^{tp} + e^{-tp} B(x)}{2} \,, \label{6}
\end{equation}
where now $p,x$ are the generators of the Heisenberg-Weil algebra,
$[p,x]=1$. (Note that the term $e^{-tp} B(x)$ in the expression for
$r$
is  understood as a multiplication of two operators but not as an
action
of the first operator on the function $B(x)$.)
We use below the following obvious relation from the theory of
Heisenberg-Weil
algebra:
$$
  e^{\mu p} F(x)= F(x+\mu) e^{\mu p}\,.
$$
Substituting (\ref{6}) into (\ref{4}) and equating to zero the
coefficients
at the operators $\exp 2tp,\exp -2tp,I$ we come to the following
system
of equations for unknown functions $A(x),B(x)$:
$$
  A(x+t)-A(x)=\tanh t\,(1-A(x+t)A(x)),\quad
  B(x+t)-B(x)=\tanh t\,(1-B(x+t)B(x))
$$
\begin{equation}
  A(x)-A(x-t)+B(x)-B(x-t)=\tanh t\,(2-A(x)B(x)-A(x-t)B(x-t))
  \label{AB}
\end{equation}
Substituting the first and the second equations (with the arguments
shifted by $-t$) into the the third one of (\ref{AB})
we obtain the selfconsistency  condition of (\ref{AB}):
\begin{equation}
  (A(x)-B(x-t))(B(x)-A(x-t))=0\,.   \label{SEL}
\end{equation}
Choosing the first possibility we have $B(x)=A(x-t)$, and the system
(\ref{AB}) becomes equivalent to a
a single equation:
$$
A(x+t)-A(x)=\tanh t\, (1-A(x+t)A(x))
$$
with the nontrivial solution:
$$
A(x)=\coth (x+F)\,.
$$
We finally obtain the explicit realization of the generators of
the two-dimensional quantum group (\ref{4}) in terms of the generators
of the Heisenberg algebra $(\hat p,\hat x)$:
\begin{equation}
s=\sinh t\hat p, \quad r=\cosh t\hat p \times \coth(\hat x+F),\quad
or \quad r=\coth(\hat x+F)\times \cosh t\hat p \label{8}
\end{equation}
where $F\equiv F(t,\hat p)$ and the two forms of the operator
$r$ corresponds to the two possible ways of resolution the
selfconsistency conditions (\ref{SEL}).

\section{The limit $t\to 0$}

To get an experience of working in the above basis let us consider
firstly the limit case $t\to 0$, assuming that the quantum algebra
is transformed into the usual $SU(2)$ algebra in this limit:
\begin{equation}
s\to ts_1,\quad r\to 1+tr_1+t^2r_2, \quad t\to 0 \label{I}
\end{equation}
(As will be clear below, it is not necessary to take into account
terms $\sim t^2$ in $s$).

Under such an assumption the principal equation (\ref{4}) (in the
leading
order with respect to $t$) takes on the form:
\begin{equation}
  [s_1,r_1]=-2r_1 \,. \label{II}
\end{equation}
On the other hand, from the definitions we have,
\begin{eqnarray*}
  r&=&\sinh t\, (T^+-T^-)+R\to 1+t(H+J^+-J^-),\\
  s&=&\sinh t(T^++T^-)\to t(J^++J^-), \qquad R\to 1+tH
\end{eqnarray*}
and indeed, Eq.\ (\ref{II}) is consistent with the last relations:
$$
  [J^++J^-,H+J^+-J^-]=-2(H+J^+-J^-)
$$
Now let us calculate $H$ from (\ref{5}). We obtain in the limit $t\to
0$,
$$
 r_1 R\to  r_1 + t\,\left({1\over 2}(r_1^2-s_1^2)-s_1+{\cal
 C}\right)\,.
$$
We thus recover the following realization of
$A_1$ algebra in the ``$L_x$'' basis:
\begin{equation}
  s_1=J^++J^-,\quad r_1=H+J^+-J^-,\quad
  r_1 H= {1\over 2}(r_1^2-s_1^2)-s_1+ {\cal C}\,.  \label{III}
\end{equation}

\section{Rising and lowering operators in the ``$J_x$''  basis}

If we assume that in the ``$L_x$'' basis the operator $s$ is diagonal,
then, in correspondence with the result of \cite{1},  its spectrum is:
$$
\hat s \ve{M}=\sinh(tM) \ve{M} .
$$
The fact of possible existence of rising (lowering) operators
$\Theta^{\pm}$ may be expressed by the equations:
\begin{equation}
  \sinh(t\hat p)\, \Theta^{\pm}=\Theta^{\pm} \sinh (t(\hat p+2))\,.
\label{RL}
\end{equation}
It is also assumed that the generators $\Theta^{\pm}$ can be
constructed from the
three generators of quantum algebra $Q^{\pm},R$  or $J^{\pm},H$.

The direct solution of this problem is unknown for us and
we will try to solve it by the trick similar to the one used
in resolution of the two-dimensional quantum group (\ref{6})
(see section 4).
We assume  that the following substitution  for
generators $\Theta^{\pm}$ works:
\begin{equation}
  \Theta^{\pm}=F^{\pm}(\hat p)\,
  (A^{\pm}(x)\exp t\hat p + \exp(-t\hat p) B^{\pm}(x))
\label{6'}
\end{equation}
Replacing the last expression into (\ref{RL}) we come to the
result:
\begin{equation}
 \Theta^{\pm}=F^{\pm}(\hat p) \exp ({\pm}2x+f^{\pm}(\hat p))
\label{LRF}
\end{equation}
The question remains, how these operators can be expressed in terms
of quantum algebra generators under realization of section 4.

\section{Matrix realization of the two-dimensional algebra}

As it follows from Section 5,
in the limit $t\to 0$  $r_1$ is the lower triangular
matrix if $s_1$ is the diagonal one. It is surprising that the
principal equation (\ref{4}) can be also resolved on the level
of $(2l+1)\times (2l+1)$-matrices ($l$ being integer or half-integer
positive number). Then $s$ is a diagonal matrix with matrix elements
$s_{i,j}=\delta_{i,j}\sinh [t(2l+2-2j)],\;(1\leq i,j \leq 2l+1)$
and $r$ the matrix with the matrix elements $r_{i,j}$
different from zero on the principal diagonal and below $(i\geq j)$.
The diagonal elements of $r$ are determined uniquely from
(\ref{4}) with the result:
$$
  r_{j,j}= \pm \cosh[t(2l+2-2j)]\,.
$$
If we want to have a correct limit $t\to 0$,
in correspondence with the results of
the section 5, the positive signs have to be taken.

Let us now rewrite (\ref{4}) in terms of the matrix elements
($i\,(j)$ is the number of the row (column) $j\leq i$):
$$
 [\sinh (t(2l+2-2i))-\sinh(t(2l+2-2j))]\alpha_{ij}
$$
$$
  = -\tanh t \,\left\{[\cosh(t(2l+2-2i))+
    \cosh(t(2l+2-2j))] \,\alpha_{ij}
  + \sum_{k=j+1}^{i-1} \alpha_{ik}\alpha_{kj}\right\}\,.
$$
The last equation may be rewritten in more compact form:
\begin{equation}
  2\cosh[t(2l+2-i-j)]\,\sinh[t(i-j-1)]\,\alpha_{ij}
  =\sinh t \sum_{k=j+1}^{i-1} \alpha_{ik}\alpha_{kj} \,. \label{MSR}
\end{equation}
From (\ref{MSR}) it follows immediately that
$\alpha_{i,i-1}\equiv \alpha_i$ are arbitrary c-numbers.
All other matrix elements of the lower triangular matrix
$\alpha_{ij}$ may be presented in the form:
$$
\alpha_{ij}=c_{ij}(t)\alpha_i \alpha_{i-1}.....\alpha_j
$$
where the functions $c_{ij}(t)$ are easily found by induction.
Firstly, let $j=i-2$. In this case (\ref{MSR} is equivalent to
$$
 2\cosh[t(2l+4-i)] \alpha_{ii-2}=\alpha_{i,i-1}\alpha_{i-1,i-2}\equiv
 \alpha_i\alpha_{i-1}
$$
or
$$
c_{i,i-2}={1\over 2\cosh[t(2l+4-2i)]}
$$
The next step $j=i-3$ leads to:
$$
c_{i,i-3}={1\over 2\cosh t(2l+4-2i)}\,{1\over 2\cosh t(2l+6-2i)}
$$
It is not difficult to prove by induction that
\begin{equation}
  c_{i,i-s}={1\over 2^{s-1}}\prod_{k=2}^s \cosh^{-1}[t(2l+2k-2i)] \,.
\label{RR}
\end{equation}
and
$$
\alpha_{i,i-s}=c_{i,i-s} \prod_{k=1}^s \alpha_{i+1-k}
$$
We thus have found (but not proved its uniqueness) realization of the
two-dimensional algebra (\ref{4}) by $(2l+1)\times (2l+1)$-
matrices depending on $2l$ arbitrary parameters $\alpha_i$.

In fact the above construction is invariant under similarity
transformations by the diagonal matrices. With the help of such a
transformation all different from zero parameters $\alpha_i$ may be
made equal to unity and we will have no arbitrary parameters in the
above
construction.  However, for further consideration it will be
convenient
do not fix these parameters.

The reader  may ask, with sarcasm, what is the role of Casimir
operator
in the whole construction? The answer is not trivial and unexpected.
Up to now, we have constructed the representation of (\ref{4})
algebra only, and after adding to it the operator $R$ from (\ref{5})
the representation of (\ref{2}) algebra, but not the representation
of the initial $SU_q(2)$ one, Eq.\ (\ref{1}).
  If we want to come back to the algebra (\ref{1}) it is necessary to
  choose
from all representations of (\ref{2}) algebra only those for which
operator
$R$ is invertible and $R^{-1}(t)=\exp-tH$ in agreement with its
definition.
However, $\det R$ defining by (\ref{5}) is equal exactly to $0$ under
the
matrix realization of this section. Indeed matrix $(s^2-r^2+1)$ is
just
lower triangular matrix with zeros on its main diagonal.
We thus need to come back to the equation (\ref{5}) in its initial
form:
\begin{equation}
  R K = {1\over 2}\,(s^2-r^2+1)\,,\quad
   K= \cosh t\,\left({\cal C} \sinh^2 t+(1-r)\cosh t - s \sinh t
   \right) \,.
\label{BK!}
\end{equation}
If we want to have a realization with $\det R$ different from zero,
then
it follows immediately from (\ref{BK!}  that $\det K=0$. $K$ is, in
general
case (with an arbitrary ${\cal C}$ in it),  the lower triangular
matrix
with different from zero diagonal elements. $\det K=0$ only
if one of its diagonal elements is equal to zero.
This defines
the possible value of ${\cal C}$ under which the transition from
(\ref{2}) to
(\ref{1}) is possible in general. In addition, in each of such cases
it is necessary to check whether the condition $R^{-1}(t)=R(-t)$
is satisfied.
Below we consider the simplest examples, explaining the situation.

\subsection{The case $2l+1=2$}

For this particular case, we have:
$$
s=\pmatrix{ \sinh t & 0 \cr
               0    & -\sinh t \cr},\quad
r=\pmatrix{ \cosh t & 0 \cr
            \alpha_1 & \cosh t \cr}.
$$
Substituting these expressions into definition of $K$ matrix (\ref{5})
we obtain:
$$
K=\pmatrix{{\cal C}\cosh t\sinh^2 t+\cosh^2 t-\cosh t & 0 \cr
-\alpha_1 \cosh^2 t & {\cal C}\cosh t\sinh^2 t+\cosh^2 t-\cosh t \cosh
2t \cr}
$$
$$
{s^2-r^2+1\over 2}=
\pmatrix{ 0 & 0 \cr
-\alpha_1 \cosh t & 0 \cr}
$$
Two possibilities arise: ${\cal C}\sinh^2 t=1-\cosh t$ and in this
case
equation
(\ref{BK!}) has no solution with $\det R\neq 0$.
In the second case, ${\cal C}\sinh^2 t=2\sinh {3\over 2} t
\sinh {1\over 2}t$ and the equation for determining
$$
R=\pmatrix{ a & b \cr
            c & d \cr}
$$
takes the form:
$$
\pmatrix{ a & b \cr
         c & d \cr}
\pmatrix{\cosh t (\cosh 2t-1) & 0 \cr
        -\alpha_1 \cosh^2 t & 0 \cr}=
\pmatrix{ 0 & 0 \cr
-\alpha_1 \cosh t & 0 \cr}
$$
with the general solution:
$$
R=\pmatrix{ A\alpha_1 \cosh t & A(\cosh 2t-1) \cr
            B\alpha_1 \cosh t & \cosh^{-1} t+B(\cosh 2t-1) \cr},\quad
\det R=A\alpha_1
$$
where $A,B$ arbitrary constants.

Choosing
$$
A\alpha_1=1,\quad A=(2 \sinh t)^{-1},\quad B=(2 \cosh t)^{-1},\quad
\alpha_1=2 \sinh t
$$
we obtain $R$ in the form:
$$
R=\pmatrix{ \cosh t & \sinh t \cr
            \sinh t & \cosh t \cr}
$$
which satisfies all of the conditions for the operator $R$.

\subsection{The case $2l+1=3$}

$$
s=\pmatrix{ \sinh 2t & 0 & 0 \cr
                   0 & 0 & 0 \cr
                   0 & 0 & -\sinh 2t \cr},\quad
r=\pmatrix{ \cosh 2t & 0 & 0 \cr
            \alpha_1 & 1 & 0 \cr
{\alpha_1\alpha_2\over 2} & \alpha_2 & \cosh 2t \cr}
$$
$$
K=\pmatrix{ \cosh t(\cosh 3t-\cosh t) & 0 & 0 \cr
           -\alpha_1 \cosh^2 t& \cosh t(\cosh 3t-\cosh t) & 0 \cr
-{\alpha_1\alpha_2\over 2}\cosh^2 t & -\alpha_2 \cosh^2 t & 0 \cr}
$$
$$
{s^2-r^2+1\over 2}=
\pmatrix{ 0 & 0 & 0 \cr
           -\alpha_1 \cosh^2 t & 0 & 0 \cr
-\alpha_1\alpha_2 \cosh^2 t & -\alpha_2 \cosh^2 t & 0 \cr}
$$
the general solution of the equation (\ref{BK!}) in this case has the
form:
$$
R=\pmatrix{ 
{T_1\alpha_1\alpha_2\over d}(1-{d\over 2}) & -T_1\alpha_2 & T_1 d \cr
{\alpha_1\over d}+{T_2\alpha_1\alpha_2\over d}(1-{d\over 2}) &
-T_2\alpha_2 & 
T_2 d \cr    
{\alpha_1\alpha_2\over 2d}+{T_3\alpha_1\alpha_2\over d}(1-{d\over 2}) 
& -T_3
\alpha_2 & 1+T_3 d \cr}    
$$
where $T_i$ three arbitrary c-number constants and $(1-{d\over
2})=\cosh 2t$.                   

We know that in the case of $A^1_q$ specter of this operator is $1,
\exp {\pm}
2t$. This means that $Det R=1, Tr R=2\cosh 2t+1,Tr R^2=2\cosh 4t+1$. 
This conditions allow to determine all constants $T_i$:
$$
{T_1\alpha_1\alpha_2\over d}=1,\quad T_2 \alpha_2 =1,\quad T_3
d=1+\cosh 2t.
$$
Putting $\alpha_1=\alpha_2=d$ we evaluate $R$ to the form
($x=1-{d\over 2}$:
$$
R=\pmatrix{ x & -1 & 1 \cr
          1+x & -1 & 1 \cr
        1+x^2 & -(1+x) & (2+x) \cr}
$$

The canonical transformation evaluating the last matrix to diagonal
form
is exactly overgo from the basis with diagonal operator $s$ to the
basis
where diagonal is $R$ operator. Explicit form of this transformation
was 
found in \cite{1}.

\section{Conclusions}

The rezults of the present paper may be devided into two part: pure
mathematical 
ones and their physical applications. As for the first ones, the
principal point 
is that we are able to reduce the three-dimensional $su_{q}(2)$
algebra 
to two-dimensional algebra ${s,r}$ (\ref{5}). Is it an exidental fact
or not?
The answer on this question will be possible only if the
generalization of the
present construction to the case an arbitrary quantum algebra will be
found.
We hope to take part in the investigation of this question in further 
publications.

No less interesting point is physical applications of this
construction. We would 
like to say a few words about its relation to a quantum optics
problems. Typical
optical hamiltonians describing the processas of thre and four-wave
mixing, the 
second and the third-harmonics generation, the interuction of atomic
systems with 
quantized cavity radiation field can be presented in the block
diagonal form.
Each block is finite dimensional subspace where the Hamiltonian acts
as three-
diagonal matrix ( usually, of very high dimension) whose matrix
elements depend
on the values of the corresponding integral of motions. The complete
solution
of the quantummachanical problem would include the diagonalization of
all these 
Hamiltonians. 

We hope that the spectrum of $"J_x"$ generators ( which in its turn
can be 
considered as some exact or appropriate Hamiltonian for some optical
model) 
found in present paper for arbitrary representation of quantum
$su_{q}(2)$ 
algebra will be usefull for analitical treatment of quantum optical
processes 
mentioned above.

\noindent{Acknowledgements}

Author would like to thanks A.V.Turbiner,V.Tolstoy and Yu.F.Smirnov
for 
interesting and fruitful discussions. 

This work was done under partial support of the Russian Foundation for 
Fundamental Recearches (RFFI) GRANT-N 98-01-00330.


\begin{thebibliography}{20}

\bibitem{1} A.\ Ballesteros and S.M.\ Chumakov, On the spectrum of
a Hamiltonian defined on su${}_q$(2). {\it quant-ph/9810061}.
To be published.

\bibitem{ZET} A.N.\ Leznov,
The Gelfand-Tsetlin selection rules and representations of quantum
algebras. {\it solv-int/9804014}. Accepted for publication NL Math.
Phys

\bibitem{Eisenhart} L.P.\ Eisenhart, {\it Continuous Groups of
Transformations}, Princeton N.J.: Princeton University Press, 1933.

\end{thebibliography}
\end{document}